\documentclass{ws-procs9x6}
\usepackage{ws-toc,ws-multind}
\usepackage{epsfig}

\newcommand{\mincir}{\raise
-3.truept\hbox{\rlap{\hbox{$\sim$}}\raise4.truept\hbox{$<$}\ }}
\newcommand{\magcir}{\raise
-3.truept\hbox{\rlap{\hbox{$\sim$}}\raise4.truept\hbox{$>$}\ }}

\begin{document}
\title{COSMOLOGY FROM XMM HIGH-Z AGN CLUSTERING} 

\author{ Manolis Plionis$^{1,2}$ \& Spyros Basilakos$^3$}

\address{$^1$ Institute of Astronomy \& Astrophysics, 
National Observatory of Athens, Greece\\ 
$^2$ INAOE, Puebla, Mexico \\
$^3$ Academy of Athens,Athens, Greece}

\begin{abstract}
We review the high-redshift X-ray
selected AGN clustering, based on the XMM/2dF survey, and compare it
with other recent XMM-based results. Using the recent Hasinger et al. (2005) 
and La Franca et al. (2005) luminosity functions we find that
the spatial clustering lengths, derived using Limber's inversion
equation, are $\sim 17$ and 20 $h^{-1}$ Mpc for the soft and hard
band sources while their median redshifts are ${\bar z}\sim 1.2$ and
0.8, respectively. 
The corresponding bias factors at $z=\bar{z}$ are
$\sim 5.3\pm 0.9$ and $\sim 5.1\pm1.1$, respectively.
Within the framework of a flat
cosmological model we find that our results support a model 
with $\Omega_m \simeq 0.26$, $\sigma_8\simeq 0.75$, $h\simeq 0.72$, w$\simeq -0.9$ 
(in excellent agreement with the 3 year WMAP results). 
We also find the present day bias of X-ray AGNs to be $b_{\circ}\sim 2$.
\end{abstract}


\section{Introduction}
Active Galactic Nuclei (AGN) can be detected out to high redshifts and
thus their clustering properties can provide
information on the large scale structure, the underlying matter
distribution and the evolution with redshift of the AGN phenomenon.
Optically selected AGN surveys miss large 
numbers of dusty systems and therefore
provide a biased census of the
AGN phenomenon. X-ray surveys, are least affected by dust providing an
efficient tool for compiling uncensored AGN samples over a 
wide redshift range. 

Early studies of the X-ray AGN clustering,
using {\it Einstein} and {\it ROSAT} data, produced 
contradictory results 
(eg. \cite{BM}; \cite{VF}; \cite{Carr}; \cite{Ak}; \cite{Mu}).
Recently, there has been an effort to address this confusing issue and
determine the clustering
properties of both soft and hard X-ray selected AGNs, based on the 
new XMM and Chandra missions (eg. \cite{Y03}; \cite{Y06}; \cite{B04};
\cite{B05}, \cite{Gil}; \cite{Pu}; \cite{Gan}; \cite{Mi06}). 
Most of these studies find a relatively large correlation length for the 
high-z X-ray AGNs
(eg., \cite{B04}; \cite{B05}; \cite{Pu}). However, the recent
XMM-COSMOS survey results \cite{Mi06} provided significantly 
smaller clustering lengths, opening again the contradiction realm.


\section{The XMM/2df survey angular and spatial clustering}
The XMM-{\it Newton}/2dF survey is a shallow (2-10\,ksec per  pointing) 
survey comprising of 18 XMM-{\it
Newton} pointings equally split between a Northern and Southern
Galactic region.
Due to elevated particle background we analysed 
a total of 13 pointings. A full description of
the data reduction, source detection and flux estimation are presented
in Georgakakis et al. \cite{Ge03}, \cite{Ge04}. 

We derive the source $\log N$-$\log S$ after constructing
sensitivity maps in order to estimate the area of the survey accessible 
to point sources above a given flux limit (see \cite{B04}, \cite{B05}).
In table 1 we also provide the effective flux-limit 
of the different surveys, estimated using the corresponding area curves, by:
$S^{\rm eff}_{\rm lim} \simeq \sum \Omega_i S_{\rm lim, i} /\sum
\Omega_i$, where $\Omega_i$ is the survey solid angle of which the
flux limit is $S_{\rm lim, i}$.

We calculate the angular correlation function
using the estimator: w$(\theta)=f (N_{DD}/N_{DR})-1$, 
of which the uncertainty is: $\sigma_{w}=\sqrt{(1+{\rm w}(\theta))/N_{DR}}$,
where $N_{DD}$ and $N_{DR}$ are the number of data-data and
data-random pairs, respectively, in the interval
$[\theta-\Delta \theta,\theta+\Delta \theta]$. 
The normalization factor is $f=2 N_R/(N_D-1)$, with
$N_D$ and $N_R$ the total number of data and random points,
respectively. 
For each XMM pointing we produce 100 Monte Carlo random 
catalogues having the same number of points as the real data 
accounting also for the sensitivity variations across 
the surveyed area (for details see \cite{B04}, \cite{B05}).
We use a standard $\chi^{2}$ minimization procedure to
fit the measured  correlation function assuming a power-law
form: w$(\theta)= (\theta_{\circ} /  \theta) ^ {\gamma-1}$ and fixing
$\gamma$ to 1.8. 
The fitting is performed for
angular separations in the range 40$^{''}-1000^{''}$
The resulting raw values of $\theta_{\circ}$
are corrected for the integral constraint and the amplification bias
(for details see \cite{VF} and \cite{B05}),
although
such corrections are quite small. The final results are presented in Table 1.
\begin{figure}[t]
{\center
\mbox{\epsfxsize=9cm \epsffile{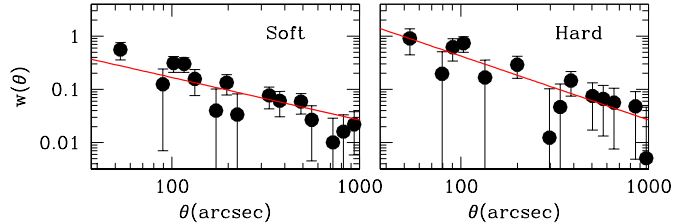}}
\caption{The XMM/2dF survey angular correlation function 
for the 2 bands considered.}
}
\end{figure}
Now, the spatial correlation function can be modeled as (eg. \cite{deZ}):
$\xi(r,z)=(r/r_{\circ})^{-\gamma}\times (1+z)^{-(3+\epsilon)}$,
where $\epsilon$ parametrizes the clustering evolution model.
For $\epsilon=\gamma-3$ (ie., $\epsilon=-1.2$ for $\gamma=1.8$), 
the clustering is constant in comoving coordinates, 
a model which appears to be
 appropriate for active galaxies (eg. \cite{Kun}).
\begin{figure}[t]
{\center
\mbox{\epsfxsize=9cm \epsffile{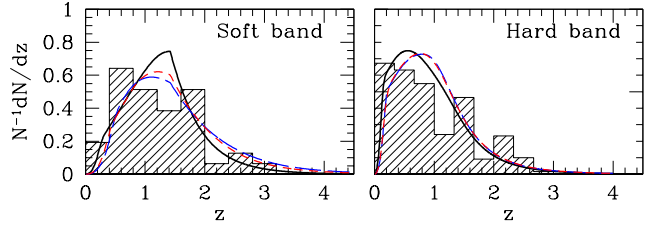}}
\caption{{\em Left Panel:} The expected $z$-distributions using the
Hasinger et al \cite{Has05} soft-band luminosity
function for the XMM/2dF survey (thick line), the XMM-COSMOS
survey (blue long-dashed line) and the XMM-ELAIS-S1 survey (red short-dashed
line). 
{\em Right Panel:} The corresponding redshift distributions
using the LDDE hard-band luminosity
function of La Franca et al. \cite{LaF05}.
The histogram in the left panel corresponds 
to the distribution of the X-ray 
sources in the ROSAT Lochman Deep Field \cite{Sch},
while that of the right panel
on limited spectroscopic and photo-z data of our XMM/2dF survey.}
}
\end{figure}

In order to invert the angular correlation function to three dimensions
we utilize Limber's integral equation (eg. \cite{Pee}). 
For a spatially flat Universe, Limber equation can be written as:
\begin{equation}\label{eq:angu}
{\rm w}(\theta)=2\frac{H_{\circ}}{c} \int_{0}^{\infty} 
\left(\frac{1}{N}\frac{{\rm d}N}{{\rm d}z} \right)^{2}E(z){\rm d}z 
\int_{0}^{\infty} \xi(r,z) {\rm d}u
\end{equation} 
where the number of objects in a survey of a solid angle $\Omega_{s}$ 
and within the shell $(z,z+{\rm d}z)$, is given by:
\begin{equation}
\frac{{\rm d}N}{{\rm d}z}=\Omega_{s}
x^{2}\phi(x)\left(\frac{c}{H_{\circ}}\right) E^{-1}(z)
\end{equation}
with $\phi(x)$ the selection function 
and $x$ the proper distance related to the redshift through (see \cite{Pee}): 
\begin{equation}
x(z)=\frac{c}{H_{\circ}} \int_{0}^{z} \frac{{\rm d}u}{E(u)}\;\; \;\; {\rm
  with}  \;\;\; 
E(z)=[\Omega_{\rm m}(1+z)^{3}+\Omega_{\Lambda}]^{1/2}\;\;,
\end{equation}
Since we do not have complete redshift information for our sources 
we estimate ${\rm d}N/{\rm d}z$ using the X-ray source luminosity function
and folding in the area curve, via the relation:  
$\phi(x)=\int_{L_{\rm min}(z)}^{\infty} \Phi(L_{x},z) {\rm d}L$,
where $\Phi(L_{x},z)$ is the 
luminosity dependent density evolution luminosity (LDDE)
function. In the present analysis we use that of Hasinger et al. \cite{Has05}
for the soft-band and that of La Franca et al. \cite{LaF05} for the hard-band.

In Fig. 2 we present the expected redshift distributions of the
soft and hard X-ray sources for all three recent XMM based surveys
together with the histogram of some
limited spectroscopic and photo-z data (see caption for details). It
is evident that all three surveys trace similar depths although the
XMM-COSMO and XMM-ELAIS surveys have a slightly larger contribution from
$z\magcir 2$ in the soft-band.
The predicted median redshift of the three XMM surveys
for the soft and hard sources are shown in table 1.

Then the inversion of eq.(2), using the LDDE luminosity function,
$\epsilon=-1.2$, the concordance cosmological model, and integrating
out to $z=4.5$, provides a
correlation length of $r_{\circ}\simeq 17\pm 2.2 \;h^{-1}$ Mpc
and $\simeq 20\pm 3.7 \;h^{-1}$ Mpc, for the
soft and hard bands, respectively (the slight differences 
with respect to the Basilakos et al. \cite{B04}, \cite{B05} results 
are due to the
different luminosity functions used and a better treatment of the errors).
These results are in good agreement with the XMM-ELAIS results \cite{Pu}
but significantly larger than the XMM-COSMOS 
\cite{Mi06} results (see table 1).

If we assume that all three XMM based surveys provide correct results,
and that their differences are due to the cosmic variance, we can volume
weight them to obtain an average estimate of the clustering
length of the XMM point sources. The relative volume weights for the 
XMM-COSMOS, XMM-ELAIS and XMM/2dF 
surveys are 1:0.9687:0.8527 and 1:0.9947:0.701 for the soft and hard bands,
respectively (we have not corrected for the slight differences of the 
hard-bands used in the three surveys).
Doing so we obtain:
$\langle r_{\circ}\rangle \simeq 13.2$ and $16.0 \; h^{-1}$ Mpc, for the
soft and hard bands, respectively.
These large correlation lengths are comparable to those of Extremely 
Red Objects \cite{Geo05}, of luminous radio  sources (\cite{Roc}; \cite{Over}; 
\cite{Rott}) and of bright distant red galaxies \cite{Fou}.
They are however, significantly larger than those 
derived from optical AGN surveys: $r_{\circ} \simeq 5.4 - 8.6 \; h^{-1}$ Mpc
(eg. \cite{Cr96}; \cite{LaF98}; \cite{Cr02};
\cite{Gra}; \cite{Por}; \cite{Wa}). 
\begin{table}[t]
\tbl{Correlation function analysis results with their 
$1\sigma$ 
uncertainties for the different
  surveys and X-ray bands. The correlation lengths assume a slope
  $\gamma=1.8$.}
{\scriptsize\begin{tabular}{@{}ccccccc@{}} \toprule
Band (keV)& Survey & \# & $S^{\rm eff}_{\rm lim}$ 
& $\theta_{\circ}$($''$)& $r_{\circ} \; (h^{-1}$ Mpc) & $\bar{z}$ \\
\colrule
 0.5-2 &  XMM/2dF & 432& 1.3 $\times 10^{-14}$ & $10.4 \pm 3$& 
$17 \pm 2.2$ &1.25\\
 0.5-2 & XMM-COSMOS &1037& 1.9 $\times 10^{-15}$ & $ 1.9 \pm 0.3$&
 $9.3 \pm 0.7$ &1.33\\
 0.5-2 & XMM-ELAIS& 395& 4.5 $\times 10^{-15}$ & $ 5.2 \pm 3.8$&
 $13.8 \pm 4.5$ &1.32\\ 
 2-8   &  XMM/2dF & 177& 4.6 $\times 10^{-14}$ & $ 22.2  \pm 9$  &
 $20 \pm 3.7$ & 0.80\\
 4.5-10& XMM-COSMOS & 151& 1.6 $\times 10^{-14}$ & $  6  \pm 2$  &
 $12 \pm 1.8$ &  0.89   \\
 2-10  & XMM-ELAIS& 205& 1.8 $\times 10^{-14}$ & $12.8 \pm 7.8$&
 $17 \pm 4.6$ & 0.88 \\\botrule
\end{tabular}} 
\end{table}

\section{Cosmological Constraints}
It is well known \cite{Kai} that according to
linear biasing the correlation function of the mass-tracer 
($\xi_{\rm obj}$) and dark-matter one ($\xi_{\rm DM}$), are related by:
\begin{equation}
\xi_{\rm obj}(r,z)=b^{2}(z) \xi_{\rm DM}(r,z) \;\;,
\end{equation}
where $b(z)$ is the bias evolution function. In this study 
we use the bias model of Basilakos \& Plionis \cite{BP01}, \cite{BP03}.
We quantify the underlying matter distribution clustering 
by presenting the spatial correlation function of the mass 
$\xi_{\rm DM}(r,z)$ 
as the Fourier transform of the 
spatial power spectrum $P(k)$:
\begin{equation}
\label{eq:spat1}
\xi_{\rm DM}(r,z)=\frac{(1+z)^{-(3+\epsilon)}}{2\pi^{2}}
\int_{0}^{\infty} k^{2}P(k) 
\frac{{\rm sin}(kr)}{kr}{\rm d}k \;\;,
\end{equation}
where $k$ is the comoving wavenumber and
$\epsilon=-1.2$, according to the constant in comoving coordinates
clustering evolution model.
As for the power spectrum, we consider that of CDM models, 
where $P(k)=P_{0} k^{n}T^{2}(k)$ with
scale-invariant ($n=1$) primeval inflationary fluctuations. 
In particular, we use the transfer function parameterization as in
\cite{Bard}, with the corrections given approximately
by Sugiyama \cite{Sug}.
Note that we also use the
non-linear corrections introduced by Peacock \& Dodds \cite{Pea}.
We have chosen to use either the standard normalization
given by: $\sigma_8 \simeq 
0.5 \Omega_{\rm m}^{-\gamma}$ with $\gamma \simeq 0.21-0.22 {\rm w} +0.33
\Omega_{\rm m}$ \cite{WS98},
or to leave $\sigma_8$ a free parameter.

Firstly, using equations (1), (4) and (5) and evaluating at $z=\bar{z}$ we derive 
the bias factor of our sources at the corresponding redshift. We find 
$b(\bar{z})\simeq 5.1\pm1.1$ and $\simeq 5.3 \pm 0.9$ for the hard and soft bands
respectively, which are significantly smaller than the values attributed to our work in 
Miyaji et al. \cite{Mi06}.

Next, we utilize a $\chi^{2}$ 
likelihood procedure to compare the measured 
XMM soft source angular correlation function  
with the prediction of different spatially flat cosmological models
(see \cite{BP05}, \cite{BP06}).
In particular, we define the likelihood estimator
as:
${\cal L}^{\rm AGN}({\bf c})\propto {\rm exp}[-\chi^{2}_{\rm AGN}({\bf c})/2]$
with:
$\chi^{2}_{\rm AGN}({\bf c})=\sum_{i=1}^{n} \left[w_{\rm th}
(\theta_{i},{\bf c})-w_{\rm obs}(\theta_{i})/\sigma_{i} \right]^{2}$, 
where ${\bf c}$ is a vector containing the cosmological 
parameters that we want to fit and $\sigma_{i}$ the observed angular 
correlation function uncertainty. 
We assume a flat ($\Omega_{\rm tot}=1$) cosmology 
with primordial adiabatic fluctuations and baryonic
density of $\Omega_{\rm b} h^{2}\simeq 0.022$ 
(eg. \cite{Kir}; \cite{Sper}).
In this case the corresponding vector
is ${\bf c}\equiv (\Omega_{\rm m},{\rm w},\sigma_8,h,b_{\circ})$
and we densely sample the various parameters.
Note that in order to investigate possible equations of state,
we allow the parameter $w$ to take values below -1, corresponding
to the so called {\em phantom} cosmologies (eg. \cite{Cald}).

The resulting best fit parameters for $\epsilon=-1.2$ and the Miyaji et al
\cite{Mi00} luminosity function
are presented in Table 2. In the first two rows we
present results based on the traditional \cite{WS98}
$\sigma_8$ normalization. Note that our estimate of the Hubble
parameter $h$ 
is in very good agreement with those derived 
by the HST key project \cite{Free}.
In the last two rows of Table 2 we leave 
$\sigma_8$ free but fix the Hubble constant to $h=0.72$.  
In this case our fit provides a $\sigma_8$ value which is in 
excellent agreement with the recent 3-years WMAP results \cite{Sper}
Therefore, 
allowing values $w<-1$ we can derive a 
$(\Omega_{\rm m},\sigma_{8})$ relation, a good fit of which is
provided by :
\begin{equation} 
\sigma_{8}=0.34 (\pm 0.01) \; \Omega_{\rm m}^{-\gamma(\Omega_{\rm m},w)}
\end{equation}
with $\gamma(\Omega_{\rm m},w)=0.22 (\pm 0.04)-0.40 (\pm 0.05)w-0.052 (\pm
0.040)\Omega_{\rm m}$.
Note that $w$ is
degenerate, within the $1\sigma$ uncertainty,
with respect to $\Omega_{\rm m}$.
Therefore, in order to put further constraints on 
$w$ we additionally use a sample of 172 supernovae SNIa of Tonry et al.
\cite{Ton}.

\begin{table}[t]
\tbl{Cosmological parameters from the likelihood analysis.
Errors of the fitted parameters 
represent $1\sigma$ uncertainties. Note that the fitted parameters
correspond to results marginalized over 
the parameters that do not have errorbars, for which
we use the values indicated.}
{\scriptsize\begin{tabular}{cccccc} \toprule
Data& $\Omega_{\rm m}$& $\sigma_8$  &$w$& $h$& $b_{\circ}$ \\
\colrule
{\rm XMM} &  $0.31^{+0.16}_{-0.08}$ & 0.93  & $w=-1$ &
$0.72^{+0.02}_{-0.18}$&  $2.30^{+0.70}_{-0.20}$ \\
{\rm XMM}/{\rm SNIa}& $0.28\pm 0.02$ &0.95  & $-1.05^{+0.10}_{-0.20}$&
$0.72$& $2.30$ \\
{\rm XMM} &  $0.28 \pm 0.03$ & $0.75 \pm 0.03$  & $w=-1$ & 0.72&  $2.0^{+0.20}_{-0.25}$\\
{\rm XMM}/{\rm SNIa}& $0.26\pm 0.04$ &0.75  &
$-0.9^{+0.10}_{-0.05}$&$0.72$ & $2.0$\\\botrule
\end{tabular}}
\end{table}
 
The joined likelihood analysis is performed by 
marginalizing the X-ray clustering results 
over $\sigma_{8}$, $h$ and $b_{0}$. The vector ${\bf c}$ now becomes: 
${\bf c}\equiv(\Omega_{\rm m}, w)$. The SNIa 
likelihood function can be written as: 
${\cal L}^{\rm SNIa}({\bf c})\propto 
{\rm exp}[-\chi^{2}_{\rm SNIa}({\bf c})/2]$, 
with:
$\chi^{2}_{\rm SNIa}({\bf c})=\sum_{i=1}^{172} \left[{\rm log}
    D^{\rm th}_{\rm L}(z_{i},{\bf c})-{\rm log}D^{\rm obs}_{\rm L}(z_{i})/
\sigma_{i} \right]^{2}$,
where $D_{\rm L}(z)$ is the dimensionless luminosity
distance, $D_{\rm L}(z)=H_{\circ}(1+z)x(z)$
and $z_{i}$ is the observed redshift. 
The joint likelihood function 
peaks at: $\Omega_{\rm m}=0.26\pm 0.04$ with $w=-0.90^{+0.1}_{-0.05}$.
Using eq.(6) we find that the normalization
of the power spectrum that corresponds to these cosmological
parameters is $\sigma_{8}\simeq 0.73$, 
in excellent agreement with the recent 3-year WMAP results \cite{Sper}.

\section*{Acknowledgments}
Many thanks are due to I. Georgantopoulos and A. Georgakakis. 


\begin{thebibliography}{99}
\bibitem{Ak}Akylas, A., Georgantopoulos, I., Plionis, M., 2000, 
MNRAS, 318, 1036
\bibitem{Bard} Bardeen, J.M., Bond, J.R., Kaiser, N. \& Szalay, A.S., 1986, 
ApJ, 304, 15
\bibitem{BP01} Basilakos, S. \& Plionis, M., 2001, ApJ, 550, 522
\bibitem{BP03} Basilakos, S. \& Plionis, M., 2003, ApJ, 593, L61
\bibitem{BP05} Basilakos, S. \& Plionis, M., 2005, MNRAS, 360, L35
\bibitem{BP06} Basilakos, S. \& Plionis, M., 2006, ApJ, 650, L1
\bibitem{B04} Basilakos, S., et al.,
2004, ApJL, 607, L79
\bibitem{B05} Basilakos, S., et al.,
2005, MNRAS, 356, 183
\bibitem{BM}Boyle, B. J., \& Mo, H. J., 1993, MNRAS, 260, 925
\bibitem{Cald}Caldell, R. R., 2002, Physics Letters B, 545, 23
\bibitem{Carr} Carrera, F.J., et al.,
1998, MNRAS, 299, 229
\bibitem{Cr96}Croom, S. M., \& Shanks, T., 1996, MNRAS, 281, 893
\bibitem{Cr02}Croom, S.M., et al.,
2002, MNRAS, 335, 459
\bibitem{deZ}de Zotti, G., et al.,
1990, ApJ, 351, 22
\bibitem{Fou} Foucaud, S., et al., 2006, MNRAS, {\em submitted},
{\tt astro-ph/0606386}
\bibitem{Free}Freedman, W., L., et al., 2001, ApJ, 553, 47
\bibitem{Gan} Gandhi, P., et al., 2006, A\&A, 457, 393
\bibitem{Ge04} Georgakakis, A., et al., 2004, MNRAS, 349, 135
\bibitem{Ge03} Georgakakis, A., et al.,
2003, MNRAS, 344, 161
\bibitem{Geo05} Georgakakis, A., et al., 2005, ApJ, 620, 584
\bibitem{Gil} Gilli, R., et al. 2005, A\&A, 430, 811
\bibitem{Gra} Grazian, A., et al.,
2004, AJ, 127, 592
\bibitem{Has05}Hasinger, G., Miyaji, T., Schmidt, M., 2005, A\&A, 441, 417
\bibitem{Kai}Kaiser N., 1984, ApJ, 284, L9
\bibitem{Kir}Kirkman, D., et al.,
2003, ApJS, 149, 1
\bibitem{Kun} Kundi\'c, T., 1997, ApJ, 482, 631
\bibitem{LaF98}La Franca F., Andreani, P., Cristiani, S., 1998, ApJ, 497,
  529
\bibitem{LaF05} La Franca, F. et al., 2005, ApJ, 635, 864
\bibitem{Mi00}Miyaji, T., Hasinger, G., Schmidt, M., 2000, A\&A, 353, 25
\bibitem{Mi06}Miyaji, T., et al., 2006, {\tt astro-ph/0612369}
\bibitem{Mu}Mullis C.R., et al.,
2004, ApJ, 617, 192
\bibitem{Over} Overzier, R.A., et al.,
2003, A\&A, 405, 53
\bibitem{Pea}Peacock, A. J., \&, Dodds, S. J., 1994, MNRAS, 267, 1020
\bibitem{Pee}Peebles P.J.E., 1993. {\em Principles of Physical Cosmology}, 
Princeton Univ.
\bibitem{Por} Porciani, C., Magliocchetti, M. \& Norberg, P., 2004,
  MNRAS,
\bibitem{Pu} Puccetti, S., et al., 2006, AA {\em submitted}, {\tt
  astro-ph/0607107}
 \bibitem{Roc}Roche, N. D., Dunlop, J., Almaini, O., 2003, MNRAS, 346, 803 
\bibitem{Rott} R\"{o}ttgering, H., et al.,
2003, New Astr. Rev., 47, 309
\bibitem{Sch}Schmidt, M., et al.,  1998, A\&A, 329, 495
\bibitem{Sper}Spergel D. N., et al., ApJ, 2006, {\em submitted},
  {\tt astro-ph/0603449}
\bibitem{Sug}Sugiyama, N., 1995, ApJS, 100, 281
\bibitem{Ton}Tonry, et al. , 2003, ApJ, 594, 1
\bibitem{VF}Vikhlinin, A. \& Forman, W., 1995, ApJ, 455, 109
\bibitem{Wa} Wake, D.A. et al., 2004, ApJ, 610, L85
\bibitem{WS98}Wang, L. \& Steinhardt, P.J., 1998, ApJ, 508, 483
\bibitem{Y03} Yang, Y., et al.
2003, ApJ, 585, L85
\bibitem{Y06}Yang, Y., Mushotzky, R.F., Barger, A.J., Cowie, L.L., 2006,
  ApJ, 645, 68
\end{thebibliography}
\end{document}